\documentclass[1p,times]{elsarticle}
\usepackage{graphicx}
\usepackage{subfigure}
\usepackage{bm}
\usepackage{amssymb}
\usepackage{multirow}
\usepackage{epstopdf}

\def\simleq{\; \raise0.3ex\hbox{$<$\kern-0.75em \raise-1.1ex\hbox{$\sim$}}\; }
\def\simgeq{\; \raise0.3ex\hbox{$>$\kern-0.75em \raise-1.1ex\hbox{$\sim$}}\; }

\newcommand{\GeV}{{\rm GeV}}
\newcommand{\TeV}{{\rm TeV}}

\newcommand{\kpc}{{\rm kpc}}

\newcommand{\m}{{\rm m}}
\newcommand{\cm}{{\rm cm}}
\newcommand{\km}{{\rm km}}

\newcommand{\s}{{\rm s}}

\newcommand{\sr}{{\rm sr}}

\newcommand{\pbar}{\bar{p}}

\graphicspath{{PLOTS/}}

\begin{document}
\begin{frontmatter}

%\hspace{11cm} {DESY 09-148}

\title{Unified interpretation of cosmic-ray nuclei and antiproton recent measurements}

\author{Giuseppe Di Bernardo$^{1,2}$, Carmelo Evoli$^{3}$, Daniele Gaggero$^{1,2}$, Dario Grasso$^{2}$, Luca Maccione$^{4}$}

\address{$^{1}$Dipartimento di Fisica ``E. Fermi", Universit\`a di Pisa, Largo B. Pontecorvo, 3, I-56127 Pisa}
\address{$^{2}$INFN, Sezione di Pisa, Largo B. Pontecorvo, 3, I-56127 Pisa}
\address{$^{3}$SISSA, Via Bonomea 265, 34136 Trieste}
\address{$^{4}$DESY, Theory Group, Notkestrasse 85, D-22607 Hamburg, Germany}

%\eads{
%\mailto{giuseppe.dibernardo@pi.infn.it},
%\mailto{evoli@sissa.it}, 
%\mailto{daniele.gaggero@pi.infn.it}, 
%\mailto{dario.grasso@pi.infn.it}, 
%\mailto{luca.maccione@desy.de}
%}

\begin{abstract}
We use our numerical code, DRAGON, to study the implications of recent data on our knowledge of the propagation properties of cosmic ray nuclei 
%with energy $\gtrsim 1~\GeV/{\rm n}$ 
in the Galaxy.
We show that B/C (as well as N/O and C/O) data, including those recently taken by CREAM, and ${\bar p}/p$ data, especially including recent PAMELA results, can consistently be fitted within a unique diffusion-reacceleration model. The requirement that light nuclei and $\bar p$ data are consistently reproduced within experimental uncertainties places significant limits on the main propagation parameters. In particular, we find the allowed range of the diffusion coefficient spectral index to be $0.3 < \delta < 0.6$ at 95\% confidence level and that
Kraichnan type diffusion is favored with respect to Kolmogorov.  
While some amount of reacceleration is required, only a limited range of the Alfv\`en velocity value ($10 \simleq v_A \simleq 20~\km~\s^{-1}$) is allowed by a combined analysis of nuclear and antiproton data, which we perform here for the first time. 
%Furthermore, we reproduce all relevant data with not need to introduce any {\it ad hoc} break in the injection spectrum of primary cosmic rays and in the diffusion coefficient dependence on rigidity. 
If antiproton data are not used to constrain the propagation parameters, a larger set of models is allowed. In this case, we determine which combinations of the relevant parameters maximize and minimize the antiproton flux under the condition of still fitting light nuclei data at 95\% C.L. These models may then be used to constrain a possible extra antiproton component arising from astrophysical or exotic sources (e.g.~dark matter annihilation or decay). 
\end{abstract}


\end{frontmatter}

\section{Introduction}

The problems of origin and propagation of Cosmic Rays (CRs) in the Galaxy are long standing questions and the combination of several different observations in a wide energy range is required to understand them at least partly. 

%Much of the difficulty in attempting at answering comes ultimately from the fact that CRs are charged particles. Being charged, they do not propagate on straight lines, but their trajectories are bent by galactic magnetic fields (GMF), hence their arrival directions on Earth do not point back to their sources. Only $\gamma$-rays (or neutrinos), produced by interactions of freshly released CRs with the interstellar medium (ISM) in the vicinity of their sources, could unveil the actual CR sources. On the other hand, the ISM in which CRs propagate is a turbulent medium. CRs interacting with randomly oriented Alfv\'en waves experience diffusion and are possibly convected and reaccelerated in the ISM. Therefore, the description of their propagation from source to Earth through the ISM is a hard task, whose degree of accuracy has necessarily grown in the years to account for the large amount of available astrophysical data (see \cite{Strong:2007nh}  for a comprehensive review). 

The most realistic description of CR propagation is given by diffusion models. Two main approaches have been developed so far: analytical (or semi-analytical) diffusion models (see e.g.~\cite{Berezinsky:book} and ref.s therein), which solve the CR transport equation by assuming simplified distributions for the sources and the interstellar gas, and fully numerical diffusion models. 
Well known realizations of these two approaches are respectively the {\it two-zone model} (see e.g. \cite{Maurin:01,Maurin:02, Maurin:2002ua}) and the GALPROP package \cite{Strong:98,Strong:04,GALPROPweb,Strong:2007nh}. Recently, some of us developed a new numerical code, DRAGON (Diffusion of cosmic RAys in Galaxy modelizatiON) \cite{Evoli:2008dv}. All these models involve in general a large number of parameters which need to be fixed using several types of experimental data. 
Their knowledge is crucial not only for CR physics but also for constraining or determining the properties of an exotic galactic component from indirect measurements.

However, in spite of the strong efforts made on both observational and theoretical sides, most of these parameters are still poorly known. One of the reasons lies in the fact that best quality data on CR spectra (e.g.~the ratios of secondary to primary nuclear species) were available mainly at low energy ($E \lesssim 10 ~\GeV/{\rm n}$), where several competing physical processes (e.g.~solar modulation, convection, reacceleration) are expected to affect significantly the CR spectra by an {\it a priori} undetermined relative amount. Furthermore, the uncertainties on the spallation cross sections and their effects on the propagated CR composition are still sizable at such low energies. 

On the other hand, the interpretation of high energy ($E > 10~\GeV /{\rm n}$) CR data is, in principle, easier since in this range only spatial diffusion and spallation losses (the latter becoming less and less relevant with increasing energy) are expected to shape the CR spectra. Furthermore, other uncertainties related to the physics of solar modulation and to poorly known nuclear cross sections are reduced by considering only data at energies larger than several GeV/n. Hence, the study of high energy CR spectra allows in principle to constrain the plain diffusion properties of CR in the Galaxy, in particular the strength $D_{0}$ of the diffusion coefficient at a reference rigidity and its energy slope $\delta$, and offers a lever arm to better understand low energy effects (see \cite{Castellina:2005ub} for an interesting discussion about this issue). 
This possibility has been precluded for long time by the scarcity of observational data. %, but 
%The scarcity of observational data has precluded this possibility for long time, but 
%Moreover, difficulties were found when trying to interpret in a plain diffusion (PD) scheme B/C and $\pbar/p$ data \cite{Moskalenko...}. [Maybe add things]
%recent measurements of the spectrum of light CR nuclei (and especially the boron to carbon ratio, B/C) up to $\sim 1~\TeV/{\rm n}$ by the CREAM balloon experiment \cite{CREAM} strongly improved the situation.

The experimental situation however improved recently when the CREAM balloon experiment measured the spectrum of light CR nuclei and especially the boron to carbon ratio (B/C) up to $\sim 1~ \TeV/{\rm n}$ \cite{CREAM}.   
Besides CR nuclear measurements, valuable complementary data were recently provided by the PAMELA satellite experiment which measured the antiproton to proton ratio up to $\sim100~ \GeV$ with unprecedented accuracy \cite{Adriani:2008zq}. 
Other valuable experimental data are expected to come from AMS-02 \cite{ams02} which will soon be installed on board of the International Space Station. 
As for other secondary nuclear species, antiprotons are produced by the spallation of primary CRs (mainly protons and Helium nuclei) in the standard scenario. Therefore, their spectrum may provide an independent and complementary check of the validity of CR propagation models and a valuable probe of an extra component which may arise, for example,  from secondary production in the CR astrophysical sources \cite{Blasi:2009bd,Blasi:2009hv} and/or from dark matter annihilation or decay (see e.g. \cite{Bergstrom:1999jc,Bergstrom:2008ag,Bertone:2008xr,Cirelli:2008pk}).  

Whether the measured secondary/primary nuclear ratios and antiproton spectra are fully compatible within the framework of a standard CR transport model is still not completely clear. 
%The compatibility of the measured secondary/primary nuclear ratios and antiproton spectra in the framework of a standard CR transport model, however, is still uncertain. 
Indeed, while a discrepancy between the parameters allowing to reproduce the B/C and the ${\bar p}/p$ was claimed in \cite{Moskalenko:2001ya}  (see also \cite{Strong:2007nh}), a good concordance was found in other analyses \cite{Bergstrom:1999jc,Donato:01}. 
Furthermore, even the interpretation of nuclear data alone is still confused: analyses based on the leaky-box and semi-analytical diffusion models favor values of $\delta$ significantly larger than the ones found with the numerical GALPROP package.  The comparison of such results is not straightforward due to a number of different assumptions. Hence, an independent analysis accounting for most recent available data is timely. 

In this work we use DRAGON \cite{Evoli:2008dv} to constrain the main diffusion parameters against updated experimental data in the energy range $1 \lesssim E \lesssim 10^3~\GeV/{\rm n}$. This code reproduces the results of the well known GALPROP under the same conditions. Furthermore, it allows to test the effects of a spatially varying diffusion coefficient. Here we use the optimized and updated version of this code, which now accounts for ionization and Coulomb energy losses, diffusive reacceleration and convection, and exploits the performances of modern computer clusters to scan a rather large range of parameters under realistic physical conditions.   
These upgrades allow to constrain the diffusion coefficient normalization and spectral index, as well as the Alfv\`en velocity $v_A$, with unprecedented accuracy by means of a statistical analysis of the agreement between model predictions and CR data including recent nuclear and antiproton data. 

In the following we will present the results of this analysis. In Sec.~\ref{sec:code} we briefly review the framework of CR propagation we adopt. In Sec.~\ref{sec:analysis} we describe our analysis and constrain the diffusion parameters. In the same section we also discuss how much the secondary antiproton spectrum is allowed to vary under the request that the predicted B/C, N/O and C/O ratios are compatible with the experimental data.
In Sec.\ref{sec:le_model} we introduce an effective diffusion-reacceleration model which allows to match all relevant experimental data down to $E \sim 0.1~\GeV/{\rm n}$. Finally in Sec.s~\ref{sec:discussion} and \ref{sec:conclusions} we compare them with results from other groups and discuss differences and implications for exotic source searches. Section \ref{sec:conclusions} is further devoted to our final remarks and conclusions.

\section{The CR propagation framework}
\label{sec:code}

Galactic CRs propagate diffusively in the irregular component of the Galactic magnetic field undergoing nuclear interactions with the gas present in the InterStellar Medium (ISM). Similarly to previous treatments, we assume here that CR Galactic source, magnetic field and gas distributions can be approximated to be cylindrically symmetric. 
Under these conditions, and in the energy range we are interested in, CR propagation of stable nuclei obeys the well known transport equation (Ginzburg and Syrovatskii \cite{Ginzburg:64}) 
\begin{eqnarray}
\label{eq:diffusion_equation}
\frac{\partial N^i}{\partial t} &-& {\bm \nabla}\cdot \left( D\,{\bm \nabla}
-\bm{v}_{c}\right)N^{i} + \frac{\partial}{\partial p} \left(\dot{p}-\frac{p}{3}\bm{\nabla}\cdot\bm{v}_{c}\right) N^i -\frac{\partial}{\partial p} p^2 D_{pp}
\frac{\partial}{\partial p} \frac{N^i}{p^2} =  \nonumber \\
&=&  Q^{i}(p,r,z) + \sum_{j>i}c\beta n_{\rm gas}(r,z)
\sigma_{ji}N^{j} -  c\beta n_{\rm gas}\sigma_{\rm in}(E_{k})N^{i}\;.
\end{eqnarray}
Here $N^i(p,r,z)$ is the number density of the $i$-th atomic species; $p$ is its momentum; $\beta$ its velocity in units of the speed of light $c$; $\sigma_{in}$ is the total inelastic cross section onto the ISM gas, whose density is $n_{\rm gas}$; $\sigma_{ij}$ is the production cross-section of a nuclear species $j$ by the fragmentation of the $i$-th one; $D$ is the spatial diffusion coefficient; $\bm{v}_{c}$ is the convection velocity; $\beta$ is the particle speed in units of $c$.
 The last term on the l.h.s. of Eq. (\ref{eq:diffusion_equation}) describes diffusive reacceleration of CR in the turbulent galactic magnetic field. In 
agreement to the quasi-linear theory we assume the diffusion coefficient in momentum space $D_{pp}$ to be related to the spatial diffusion coefficient by the relationship (see e.g.~\cite{Berezinsky:book}) $\displaystyle D_{pp} = \frac{4}{3 \delta (4 - \delta^2)(4 - \delta)} v_A^2~p^2 / D$ where  $v_A$ is the Alfv\`en velocity. Here we assume that diffusive reacceleration takes place in the entire diffusive halo. 

Although DRAGON allows to account also for CR convection, we neglect this effect in the present analysis showing {\it a posteriori} that it is not necessary to consistently describe all the available data above $1~\GeV/{\rm n}$ (see Sec.~\ref{sec:discussion}). Hence in the following we will set $v_{c} = 0$.
By this we do not mean that CR data implies that the physical value of $v_{c}$ is actually vanishing but only that an effective description of their propagation is possible even if convection is disregarded (see the discussion at the end of Sec.\ref{sec:discussion}).

DRAGON~\cite{Evoli:2008dv} solves Eq.~(\ref{eq:diffusion_equation}) numerically in the stationary limit $\partial N_{i}/\partial t = 0$ 
by imposing the following boundary conditions: $N(p,R_{\rm max},z) = N(p,r,z_{\rm min}) = N(p,r,z_{\rm max}) = 0$, corresponding to free escape of CRs at the outer limit of the Galaxy; a symmetry condition on the axis $r = 0$, $N(p,0+\epsilon,z) = N(p,0-\epsilon,z)$ ($\epsilon \ll 1$), due to the assumed cylindrically symmetric setup; a null flux condition $\partial N/\partial p = 0$ on the momentum boundaries, which stems for the fact that particles at null momentum should not lose momentum anymore. Even though the presence of reacceleration can effectively invalidate this condition, by producing a net flux from low to high momenta, we remark that this can affect only the part of the spectrum close to the momentum boundary. For this reason, we adopt an energy grid whose extrema are well below and well above the minimal and maximal energy of the data set we consider. In such a way, our results are in fact independent of the momentum boundary conditions we impose. 
The spatial limits of our simulation box are defined by $R_{\rm max} = 20~\kpc$ and $z_{\rm max} = -z_{\rm min}$. We start the spallation routine from $Z = 16$, having verified that the effect of heavier nuclei on the results of the present analysis is negligible when compared to other uncertainties, being below the 1\% level. 

We briefly recall below the main assumptions we make for the terms appearing in Eq.~(\ref{eq:diffusion_equation}).
%%%%%%%%%%%%%%%%%%%%%%%%%%%%%%%%%%%%%%
%\begin{description}
%%%%%%%%%%%%%%%%%%%%%%%%%%%%%%%%%%%%%%
%\item[Spatial diffusion coefficient] 

\subsection{Spatial diffusion coefficient}
%We assume cylindrical symmetry and that the regular magnetic field is azimuthally oriented $({\bf B_0} = B_\phi(r,z)\,\hat{\bm{\phi}})$. Under these conditions CR diffusion out of the Galaxy takes place only perpendicularly to ${\bf B_0}$. Therefore $D$ represents in fact the perpendicular diffusion coefficient $D_\perp$. 
The dependence of the diffusion coefficient $D$ on the particle rigidity $\rho$ and on the distance from the Galactic plane $z$ is taken to be (here we assume $D$ to be cylindrically symmetric and independent on the Galactocentric radius $r$)
\begin{equation}
\label{eq:diff_coeff}
 D(\rho, z) = D_0 ~\beta^\eta \left(\frac{\rho}{\rho_0}\right)^\delta\ ~ \exp\left\{|z|/z_t \right\}\;,
 \end{equation}
 where $\beta$ is the particle velocity in units of the speed of light $c$. 
As shown in \cite{Evoli:2008dv},  a vertically growing $D$ is physically more realistic than a uniform one and allows to get a more regular behavior of the CR density at the vertical boundaries of the propagation halo with respect to the case of uniform diffusion. As far as the analysis discussed in this paper is concerned, however, the substitution of such a profile with a vertically uniform $D$ only requires a change of the normalization factor $D_0$. 
Generally, the value $\eta =1$ is adopted in the related literature. This parameter, however, is not directly constrained by independent observations 
and other values have been recently considered (see e.g. \cite{Maurin:09}).
We neglect here a possible dependence on the radial coordinate $r$, which was considered also in \cite{Evoli:2008dv}.
We always set $z_{\rm max} = 2\times z_{t}$ in Eq.~(\ref{eq:diff_coeff}) to avoid border effects, and $\rho_{0} = 3~{\rm GV}$ in the following.
Finally, we assume no break in the power-law dependence of $D$ on rigidity, and we checked that our results do not depend on the choice of $z_{\rm max}$, but only on $z_{t}$, which then acts as the effective vertical size of the diffusive halo.  
%%%%%%%%%%%%%%%%%%%%%%%%%%%%%%%%%%%%%%  
%\item[CR sources] 

\subsection{Cosmic ray sources}
For the source term we assume the general form 
\begin{equation}
Q_{i}(E_{k},r,z) =  f_S(r,z)\  q^{i}_{0}\ \left(\frac{\rho(E_{k})}{\rho_0}\right)^{- \alpha_i} \;,
\end{equation}
and impose the normalization condition $f_{S}(r_{\odot},z_{\odot}) = 1$.
We assume $f_S(r,z)$ to trace the SNR distribution as modeled in \cite{Ferriere:01} on the basis of pulsar and progenitor star surveys \cite{Evoli:2007iy}. 
This is slightly different from the radial distributions adopted in \cite{Strong:04b} and in \cite{Maurin:01,Maurin:2002ua} which are based on pulsar surveys only.
Two-zone models assume a step like dependence of $f_S(r,z)$ as function of $z$, being 1 in the Galactic disk ($|z| < z_d$) and 0 outside.    
%  Almost coincident results are obtained if the same distribution is adopted.  
%In the galactic disk such a distribution is similar to that adopted in \cite{Strong:04b}, but shows an excess in the Galactic Bulge due to the contribution of type-Ia SNe, not accounted for in \cite{Strong:04b}. %Both distributions are significantly more peaked than those empirically determined \cite{Mattox:96,Strong:98} by matching the $\gamma$-ray longitude profile measured by EGRET \cite{Hunter:97}. 
%Although DRAGON would allow us to use different spectral indexes $\alpha_{i}$ for different nuclei, we do not exploit this possibility here.
%Even though allows to consider different  power-law indexes $\alpha_i$ for the different nuclear species, in this work we only consider the same $\alpha_{i}\equiv\alpha$ for all species, when not differently stated. 
For each value of $\delta$ in Eq.~(\ref{eq:diff_coeff}) we fix $\alpha_i$ by requiring that at very high energy ($E_k \gg 100~\GeV$/n) the equality $\alpha_i + \delta = \gamma_i$ holds, as expected in a plain diffusion regime. Indeed, at such high energies reacceleration and spallation processes are irrelevant.
Here we adopt the same spectral index ($\gamma_i = \gamma$, hence $\alpha_i = \alpha$)  for all nuclei as indicated by recent experimental results \cite{CREAM3,Boyle:2008ut,Ave:2008uw}. 

The low energy behavior of $Q$ is quite uncertain and several different dependencies of $Q$ on the velocity $\beta$ have been considered (see e.g.~\cite{Maurin:01}).  In the energy range explored in this work, however, different choices of such behavior have negligible effects. This strengthens further the importance of relying on high energy data to reduce systematic uncertainties.

The injection abundances $q^i_0$ are tuned so that the propagated, and modulated, spectra of primary species fit the observed ones.  Here we choose to normalize the source spectra of Oxygen and heavier nuclides to reproduce the observed spectra in CRs at $E \sim 100~\GeV$/n. On the other hand, Carbon and Nitrogen (which, together with Oxygen mostly affect the B/C) injection abundances (with respect to Oxygen) are free parameters, over which we marginalize our statistical variables in our analysis, in a way which we will describe in Section \ref{sec:analysis}. Our data basis for Oxygen and heavier nuclei is constituted by ACE/CRIS data \cite{ACE}. For B, C and N, besides CREAM's,  we use experimental data provided by the HEAO-3 \cite{HEAO-3} and CRN \cite{CRN} satellite-based experiments.  HEAO-3 B/C data are nicely confirmed by a recent preliminary analysis of AMS-01 data \cite{AMS1_BC} which, however, we do not use in this work. 
 
%A detailed analysis, accounting for data over the entire rigidity range considered here, will be performed to fix the C, N and O relative source ratios (see below) as these quantities mostly affect the B/C. Rather, the normalization of the source spectra of Oxygen and heavier nuclides is tuned to reproduce the observed spectra in CRs at $E \sim 100~\GeV$/n. 
We verified {\it a posteriori} that the observed Oxygen spectrum (see below), as well as the subFe/Fe ratios\footnote{ To compute these ratios, of course we extended our numerical simulations up to $Z=28$.}, are reasonably reproduced by our best-fit model. 

For the primary proton local interstellar spectrum (LIS) we adopt $J_p = 1.6 \times 10^4\ (E_k/1~\GeV)^{-2.73}\\  ~(\m^2~\s~ \sr~ \GeV)^{-1}$ as measured by BESS during the 1998 flight \cite{Sanuki:2000wh}. This spectrum also provides an excellent fit to AMS-01 \cite{AMS01} data and, as we will show below, also to preliminary PAMELA proton spectrum data \cite{PAMELA:proton}. 

What is most important here, however, is that we assume no spectral breaks in the source spectrum of all nuclear species. 
As we will discuss in Sec.~\ref{sec:discussion}  this point is crucial to understand the difference between our results and those of some previous works. 

%%%%%%%%%%%%%%%%%%%%%%%%%%%%%%%%%%%%%%
%\item[Nuclear cross sections] 
\subsection{Nuclear cross sections}
The spallation cross sections and the spallation network are based on a compilation of experimental data  (when present) and semi-empirical energy dependent interpolation formulas as provided e.g.~in \cite{Letaw:83,Webber:90,Silbeberg} (see also GALPROP, \cite{GALPROPweb} and references therein, from which data and some related routines have been obtained and included in DRAGON as an external library).  

For antiprotons, the main processes responsible for their production are $p - p_{\rm gas}$, $p - {\rm He}_{\rm gas}$, ${\rm He} -  p_{\rm gas}$ and 
${\rm He} - {\rm He}_{\rm gas}$ reactions, plus a negligible contribution from other nuclei. Similarly to \cite{Moskalenko:2001ya,Donato:01} we adopt the ${\bar p}$ production cross-section calculated using the parametrization given in Tan \& Ng \cite{Tan:1982nc}. 
%We account for the contribution of heavier nuclei in the CRs and the ISM by using the effective correction function determined by Simon {\it et al.}~\cite{Simon:98} with the MonteCarlo model DTUNUC. 
Inelastic scattering, annihilation and tertiary ${\bar p}$ (antiprotons which have been inelastically scattered) are treated as in \cite{Moskalenko:2001ya}. 

In order to test the possible dependence of our results on systematical uncertainties on those cross sections, we performed several DRAGON runs using also a different set of nuclear cross sections as determined in \cite{Webber:03} (see Sec.\ref{sec:discussion}). 

%{\sl An important point to be noticed here is that although the overall uncertainties on the nuclear cross sections are large, several of them are reduced above few hundreds MeV/n and, most importantly, they are less and less relevant with increasing energy, due to the increased relative role of diffusion in shaping CR spectra. This is one of the main reasons to consider only CRs with $E > 1~\GeV{\rm /n}$ and to repeat the analysis for larger values of $E_{\rm min}$ as we do in this work for the first time (see below).}     

%%%%%%%%%%%%%%%%%%%%%%%%%%%%%%%%%%%%%%
%\item[Target gas] 
\subsection{Target gas}
The ISM gas is composed mainly by molecular, atomic and ionized hydrogen (respectively, H$_2$, HI and HII). 
%Although more realistic distributions are known, for $r>2~\kpc$ 
Here we adopt the same distributions as in \cite{Strong:98,Evoli:2008dv}. 
We checked that other possible choices do not affect significantly our final results.  
%, for essentially two reasons. First of all, since CRs propagate for million years in the Galaxy, in the stationary limit they just probe a smoothed, mean gas distribution. Secondly, we can have a more direct comparison with GALPROP results.

%However, in the central region of the Galaxy, where GALPROP assumes an interpolated density, we use the the H$_2$ and HI distributions as modeled in \cite{Ferriere:07}. While the flux and composition of charged CR reaching the Earth are not sensitive to the central gas distribution, this choice allows us to better model the $\gamma$-ray emission in the Galactic Centre (GC) region, as we will discuss in more details in sec.~\ref{sec:gamma}. 
Following \cite{Asplund:2004eu} we take the He/H numerical fraction in the ISM to be 0.11. We neglect heavier nuclear species.

%\item[Solar modulation]
\subsection{Solar modulation}
We describe the effect of solar modulation on CR spectra by exploiting the widely used force-free approximation \cite{Gleeson&Axford}, prescribing that the modulated spectrum $J(E_k,Z,A)$  of a CR species is given, with respect to the Local Interstellar Spectrum (LIS) $J_{\rm LIS}(E_k,Z,A)$, by 
\begin{equation}
\label{eq:modulation}
J(E_k, Z, A) =  \frac{ (E_k + m)^2 - m^2}{\left(E_k + m + \frac{Z|e|}{A} \Phi \right)^2 - m^2}\  J_{\rm LIS}(E_k  + \frac{Z|e|}{A} \Phi, Z, A)\;,
\end{equation}
where $m$ is the nucleon mass and $\Phi$ is the so called modulation potential. This potential is known to change with the solar activity with a period of 11 years. 
%The effects of modulation are significant only for $E_k  \lesssim | Ze | /A\Phi$.
It must be stressed that the potential $\Phi$ is not a model independent quantity. Rather, for each propagation model it should be obtained by fitting the CR spectra at low energy. The possibility of restricting our analysis to $E_{k} > 1~\GeV/{\rm n}$ will reduce the systematic uncertainties associated to this unknown.  Above $1~\GeV/{\rm n}$ the effects of modulation on the secondary/primary CR ratios used in our analysis 
are tiny and can safely be accounted for by means of the simple force free approximation. 
 
For protons and antiprotons we use a potential which allows to match BESS98 \cite{Sanuki:2000wh}, AMS-01\cite{AMS01} and PAMELA \cite{PAMELA:proton} proton data even well below 1 GeV/n (see Fig.~\ref{fig:protons}). Indeed all these experiments took their data in a period with almost the same, almost minimal, solar activity.  Although a more complicated and realistic treatment of solar modulation, accounting for charge dependent effects, and the 22 year cycle change of polarity associated to solar effects, might be needed when dealing with ${\bar p}/p$ ratios for $E_{k} \simleq 1~\GeV/{\rm n}$
(see e.g.~\cite{Bieber:1999dn}), we decide to work in the framework of the force-free field approximation and show a posteriori that the data considered in our analysis can naturally be described in that framework. 

%{\bf For secondary/primary light nuclei ratios we use $\Phi = 650~{\rm MV}$  which allows us to reproduce Oxygen taken by ACE/CRIS \cite{ACE}  near solar minimum.}
%We verified that changing $\Phi$ within a reasonable interval do not affect significantly the final results of our analysis.  
%\end{description}

\section{Analysis and results}
\label{sec:analysis}

Our goal is to constrain the main propagation parameters $\delta$,  $D_{0}$,  $z_{t}$ and $v_A$ entering Eq.~(\ref{eq:diff_coeff}).  
%Although our code allows to account for the effects of a convective wind,  we verified that this is not required to consistently interpret all data for $E > 1~\GeV/{\rm n}$ and consequently assume a vanishing convective velocity in our analysis. 
To this aim, we compare to experimental data our prediction for the following physical quantities: the B/C, N/O, C/O ratios for $1 <  E_{k} <  10^3~\GeV/{\rm n}$ and the 
 $\bar{p}/p$ ratio for  $1 < E_{k} < 10^2~\GeV/{\rm n}$.  We will check {\it a posteriori} that also the Oxygen, proton and antiproton absolute spectra are correctly reproduced by our preferred models. 
In order to test the relevance of low energy physics on our constraints of the diffusion-reacceleration parameters, we perform our analysis for three different values of the minimal energy $E_{\rm min}$. We will then motivate the choice of the most suitable value of $E_{\rm min}$.  

%These are primary/primary and secondary/primary ratios.  In particular, the spectrum of secondary/primary ratios allows us to infer informations directly on $\delta$ \cite{Berezinsky:book}. 
%Nuclei heavier than oxygen play a minor role here and their abundances are fixed so to reproduce the observed abundance of those CR elements within they large uncertainties.  

As long as the propagation halo scale height is allowed to vary within the range $2 \lesssim z_t \lesssim 6~\kpc$ (which is what we assume here),
$D_{0}$ and $z_{t}$ are practically degenerate so that our results depend only on the ratio $D_{0}/z_{t}$. Throughout this paper we will always express this quantity in units of  $10^{28}~\cm^2~\s^{-1}~\kpc^{-1}$.  
We verified {\it a posteriori} that for this range of $z_t$ values, the predicted $^{10}$Be/$^{9}$Be ratio, which constrains the CR propagation time hence 
the vertical scale height of the propagation region \cite{Berezinsky:book} when combined with secondary/primary stable nuclei data,
is consistent with experimental data. 
%A possible way to break this degeneracy is to consider unstable to stable ratios (e.g.~$^{10}$Be/$^{9}$Be), which are known to probe the propagation time, hence 
%the vertical scale hight of the propagation region \cite{Berezinsky:book} when combined with the grammage as determined from the secondary/primary stable nuclei ratios.  Unfortunately, unstable nuclei fluxes have been reliably measured only up to few $\GeV/{\rm n}$. As a consequence the diffusion halo scale height  can be estimated only with a large uncertainty.  On this basis we find  the preferred range $3 < z_{t} < 5~\kpc$  \cite{Evoli:2008dv,Moskalenko:2001qm} {\bf VERIFICARE !}. 

\subsection{Light nuclei ratios}
\label{sec:nuclei_analysis}

\subsubsection{Method}
We already showed \cite{Evoli:2008dv} that in order to constrain correctly the propagation parameters on the basis of B/C measurements it is essential to take into proper account that the main primary parent species of Boron\footnote{ A non negligible contribution to the $^{10}$B comes from the beta decay of $^{10}$Be, which is properly accounted for in our analysis.} are also affected by propagation. This holds not only for the Nitrogen (N = $^{14}$N +  $^{15}$N), which gets a significant secondary contribution, but also for Carbon and Oxygen, since for  $E_k < 100~\GeV/{\rm n}$ their spectra are shaped by spallation losses in a propagation dependent way. Therefore, we perform our likelihood analysis in three steps: 
\begin{enumerate}
\item  for fixed values of the propagation parameters $v_A$,  $\delta$, and $D_{0}/z_{t}$ we vary the C/O and N/O source ratios to compute the $\chi^{2}\ $\footnote{Every time we refer to a $\chi^{2}$, we mean the $\chi^{2}$ divided by the number of degrees of freedom, i.e.~the so called reduced $\chi^{2}$.} (which we call  $\chi^{2}_{\rm C,N,O}$) of the propagated, and modulated, C/O and N/O ratios against experimental data in the energy range $1 < E_k < 10^3~{\GeV/{\rm n}}$;  
\item for the same fixed value of $v_A$, we finely sample the parameter space ($\delta$, $D_{0}/z_{t}$) by using, for each couple of these parameters, the C/O and N/O source ratios which minimize $\chi^{2}_{\rm C,N,O}$; for each of these realizations we compute the $\chi^{2}$ (which we call $\chi^{2}_{\rm B/C}$) for the B/C modulated ratio against data with $E > E_{\rm min}$;
\item we repeat the same analysis for several values of $v_A$ to probe the effect of diffusive reacceleration. For each value of $v_A$ we then determine the allowed ranges of $\delta$ and $D_{0}/z_{t}$ for several Confidence Levels (CL).  
 \end{enumerate}
 
In \cite{Evoli:2008dv} only items (i) and (ii) were performed, for $v_A = 0$ and without accounting for CREAM data, not yet public at that time. 

%Here, besides CREAM's,  we use experimental data provided by the HEAO-3 \cite{HEAO-3} and CRN \cite{CRN} satellite-based experiments.  HEAO-3 B/C data are nicely confirmed from a recent preliminary analysis of AMS-01 data \cite{AMS1_BC} which, however, we do not use in our work. 

The wide energy range covered by these recent data allows us to perform our analysis using three different energy intervals defined by $E_{\rm min} = 1,\,  5$ and $10~{\GeV/{\rm n}}$ respectively and by the same $E_{\rm max} = 1~\TeV/{\rm n}$.
 
As we already stated in the above, we do not account in our analysis for light nuclei and antiproton data with energy below $1~\GeV/{\rm n}$ as they are affected by poorly known low energy physics and are not necessary to constrain the high energy behavior of the diffusion coefficient, which is the main goal of this work.  In Sec. \ref{sec:le_model} we will show, however,  that 
 specific models which fit all data even below that energy can be built, adopting diffusion coefficients allowed from our analysis.
 
 % which we may have not accounted for. {\bf (Altrimenti sembrava che avessimo trascurato qualcosa deliberatamente)}
 %By iterating this procedure for several values of the pair ($D_0/z_t$, $\delta$), we sample the whole parameter space of our interest. Minimization of $\chi^{2}_{\rm B/C}$ will lead to our best fit values for ($D_0/z_t, \delta$) and the appropriate confidence regions.

%\resizebox{8cm}{!}

\subsubsection{Results}
In Tab.~\ref{tab:analysis} we report the best-fit model parameters, and the relative minimal $\chi^{2}_{\rm B/C}$'s, as determined for several values of $v_A$ and $E_{\rm min}$.

\begin{table}[tbp]
\centering
\caption{Best fit parameters, and the corresponding $\chi^2$ values resulting from comparing our model predictions with nuclear experimental data alone (B/C analysis) and with nuclear and $\pbar/p$ combined data (combined statistical analysis), as described in text. The values corresponding to $E_{\rm min} = 5~\GeV/{\rm n}$ for the combined analysis, which are used to constraint our models, are reported in bold.}
\begin{tabular}{|l|l|l|l|l|l|l|l|l|l|}
\hline  \multicolumn {2}{|c|} {   } & \multicolumn{3}{|c|}{B/C analysis}  & \multicolumn{3}{|c|}{joint analysis}  \\
\hline $v_A \,{[\rm km/s]}$ & $E_{\rm min}\, [\GeV/{\rm n}]$ & $\delta$ & 
$D_{0}/z_{t}$ & $\chi^2$  & $\delta$ & $D_{0}/z_{t}$ & $\chi^2$ \\
\hline
\multirow{3}{*}{0}        & 1   &        0.57  &       0.60  & 0.38    & 0.47 & 0.74 & 3.25 \\
                                   & 5   &        0.52  &       0.65  & 0.33    & {\bf 0.41} & {\bf 0.85} & {\bf 2.04} \\
                                   & 10 &        0.46  &       0.76  & 0.19    & 0.44 & 0.82 & 1.57 \\
\hline
\multirow{3}{*}{10}      & 1   &        0.52  &       0.68  & 0.32    & 0.49 & 0.71 & 1.47 \\
                                   & 5   &        0.49  &       0.71  & 0.28    & {\bf 0.41} & {\bf 0.85}  & {\bf 1.69} \\
                                   & 10 &        0.44  &       0.82  & 0.20    &  0.44 & 0.82 &  0.12 \\
\hline
\multirow{3}{*}{15} & 1   &  0.46 &  0.76 & 0.33    & 0.47 & 0.76 & 0.94 \\
                                   & 5   &       0.49   &       0.73  & 0.26    & {\bf 0.44} & {\bf 0.82} & {\bf 0.12} \\
                                   & 10 &       0.44   &       0.84  & 0.18    & 0.41 & 0.98 & 0.16 \\
\hline
\multirow{3}{*}{20}      & 1   &       0.41   &       0.90  & 0.47    & 0.47 & 0.79 & 2.28 \\
                                   & 5   &       0.44   &       0.84  & 0.22    & {\bf 0.44} & {\bf 0.84} & {\bf 0.85} \\
                                   & 10 &       0.44   &       0.87  & 0.20    & 0.44 & 0.85 & 0.98 \\
\hline
\multirow{3}{*}{30}      & 1   &       0.33   &       1.20  & 0.40    & 0.33 & 1.20 & 5.84 \\
                                   & 5   &       0.38   &       1.06  & 0.20    & {\bf 0.36} & {\bf 1.09} & {\bf 2.47} \\
                                   & 10 &       0.41   &       0.98  & 0.16    & 0.38 & 1.04 & 1.61 \\
\hline

\end{tabular}
\label{tab:analysis}
\end{table}
\begin{figure}[tbp]
\centering
 \includegraphics[scale=0.5]{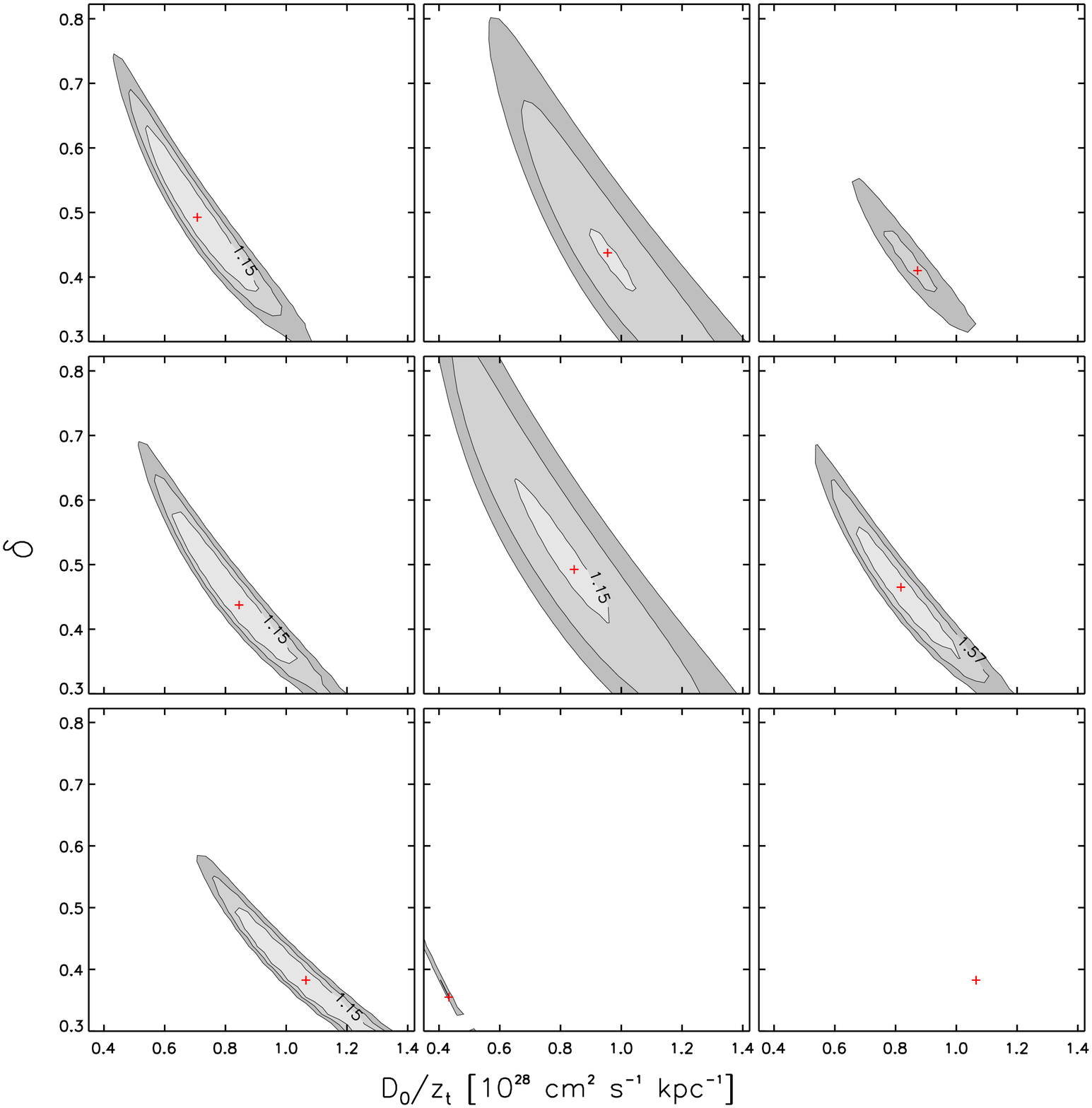}
\caption{The 68\%, 95\% and 99\% confidence level regions of DRAGON models, computed for $E_{\rm min} = 5~\GeV/{\rm n}$ are represented in the plane $(D_{0}/z_{t},\delta)$.  For the 68\% confidence level the corresponding value of the $\chi^2$ is also shown. The red crosses show the best-fit position. 
Each row corresponds to different values of the Alfv\`en velocity: $v_A = 10,20,30~\km/\s$ from top to bottom. 
Each column corresponds to different analyses: B/C (left panels), $\bar{p}/p$ (center panels) and combined (right panels).
% are represented in the plane $(D_{0}/z_{t},\delta)$  for several values of the Alfv\`en velocity $v_A = 10,15,20~\km/\s$ from top to bottom.
}
\label{fig:CL}
\end{figure}

First of all we notice that in the highest energy range ($E_{\rm min} = 10~{\GeV/{\rm n}}$) the best-fit model values of $\delta$ and $D_{0}/z_{t}$ are weakly dependent on the Alfv\`en velocity. In particular, the best fit values of $\delta$ stays in the very narrow range $0.40 \div 0.46$ varying $v_{A}$ from  $0$ to $30~\km/\s$.
%
%Especially for what concerns $\delta$, we see that for all considered choices of $v_A$,  it best-fit values vary with $v_A$ within a very narrow range 
%($0.41 \div 0.46$). 
This agrees with the common wisdom that reacceleration is almost ineffective at such high energies (see also  Fig.~\ref{fig:BC_variVA}).
%Therefore, the analysis performed for $E_{\rm min} = 10~{\GeV/{\rm n}}$ best probes indeed the actual physical values of $\delta$ and $D_{0}/z_{t}$, {\bf though with a still large statistical uncertainty}.    

The most useful results, however, are those obtained for $E_{\rm min} = 5~{\GeV/{\rm n}}$ since that threshold provides the best compromise between the two opposite requirements: 
1) to include in the analysis more experimental data and  2)  to work in an energy range where propagation is as less as possible affected by poorly known low energy physics. For example, possible charge dependent drift effects in the solar modulation (see e.g. \cite{Bieber:1999dn,Moskalenko:2001ya}) can be safely neglected in that energy range. Best fit parameters and confidence level contours obtained for that value of $E_{\rm min}$ are showed in Tab.~\ref{tab:analysis} and in Fig.~\ref{fig:CL} respectively.

From both we notice that  all considered values of $v_A$ are almost equally permitted by the B/C $\chi^2$ analysis, and that the $\delta - D_0/z_t$  allowed region slightly moves towards low $\delta$'s and large $D_0/z_t$'s as $v_A$ is increased from 0  to $30~\km/\s$.  While Kraichnan diffusion is clearly favored in the case of low values of $v_A$, Kolmogorov becomes favored, for $v_A \simgeq 30 ~\km/\s$. The choice among those model, however, is difficult in the absence of an independent estimate of $v_A$.  We will show that the antiproton/proton data break such degeneracy.  

In Fig.~\ref{fig:BC_variVA} we show the effect on the B/C ratio of varying $v_A$ keeping $\delta$ and $D_{0}/z_{t}$ fixed to the value $(0.45, 0.8)$ which will be motivated below.

\subsection{Antiprotons}
\label{sec:antip_analys}

%\subsubsection{Method}
The statistical analysis for the $\bar{p}/p$ ratio is rather simpler than the one for B/C. Indeed, the secondary $\bar p$ production depends, besides on $D_{0}/z_{t}$, $\delta$ and $v_A$, only on the source abundance ratio He/p. This last unknown quantity can be easily fixed by looking at the measured spectrum of He at Earth, which is relatively well known. %Therefore, we construct a $\chi^{2}_{\pbar/p}$ by comparing our predicted $\pbar/p$ spectrum for different values of ($D_{0}/z_{t},\delta$) to experimental observations. 
Therefore, we do not need to fit the source abundance ratio here and can directly proceed to map the $\chi^{2}_{\bar{p}/p}$  in the ($D_{0}/z_{t},~\delta$) space, for several $v_A$,  similarly to what described in items (ii) and (iii) of the previous subsection.
 
%The results of this analysis are reported in Tab. 2.  

%\subsubsection{Results}
In the second column of Fig.~\ref{fig:CL} we show the statistically allowed regions in the plane $(D_{0}/z_{t},\delta)$ for several values of $v_A$ and compare them with the corresponding regions determined from the light nuclei analysis (first column in the same figure). The allowed CL region is significantly larger than that determined from the light nuclei analysis (due to the larger experimental errors) and they overlap only for some values of the Alfv\`en velocity. 
In fact, it is remarkable that the $v_A$ varying behavior of those regions is almost opposite so that not all values of $v_A$ are allowed by a combined analysis (see Sec. \ref{sec:comb}). 

\subsection{Combined analysis and constraints on the propagation parameters}\label{sec:comb}

A combined analysis of light secondary/primary nuclei and antiproton/proton data can be performed under the working hypothesis that CR antiprotons are only of secondary origin.  

We define the combined reduced $\chi^2$ as $\displaystyle \chi^2_{\rm comb} = \frac{1}{2} \left( \chi^2_{\rm BC} + \chi^2_{\rm ap/p} \right)$.
The CL regions for several values of $v_A$ are reported in the third column of Fig.~\ref{fig:CL} and the corresponding best-fit parameters in Tab.~\ref{tab:analysis}.  
Again, here we use only data with  $E > E_{\rm min} = 5~\GeV/{\rm n}$. 

As we anticipated in the previous subsection, in general the CL region allowed by the combined analysis is smaller than the B/C one. 

Indeed, while the parameter regions constrained by the B/C and $\bar{p}/p$ data nicely overlap
for $10 \simleq v_A \simleq 20~\km~ \s^{-1}$, models outside this range do not allow a combined fit of both data sets at the required level of statistical significance (higher than 95\%). 
The fact that only a limited range of the Alfv\`en velocity values are allowed is consequence of the different behavior of the B/C and ${\bar p}/{p}$ ratios with $v_A$ due to the different spectral shapes of these ratios. This is a new and quite interesting results.  

It is reassuring to notice that the results of the analysis performed for $E_{\rm min} = 5$ and $10~\GeV/{\rm n}$ are practically coincident, which makes us confident that the combined analysis performed for $E_{\rm min} = 5~\GeV/{\rm n}$ probes already the purely diffusive CR regime.
It is also remarkable that the best fit values of $\delta$ and $D_0/z_t$ stay almost unchanged when varying $v_A$. In particular $(\delta, D_0/z_t) \simeq (0.4 - 0.45, 0.8)$  for all allowed values of $v_A = 10 - 20~\km \s^{-1}$ of the Alfv\`en velocity.  This makes us confident that the combined analysis performed for $E_{\rm min} = 5~\GeV/{\rm n}$ best probes the diffusion-reacceleration parameters . 

%were already excluded by the $\bar{p}/p$ data alone. Here we see that even for the allowed values of $v_A$,  the parameter region constrained by the light nuclei analysis narrows significantly when ${\bar p}/p$ is taken into account.

\begin{figure}[tbp]
  \centering
 \subfigure[]
 {
   \includegraphics[scale=0.4]{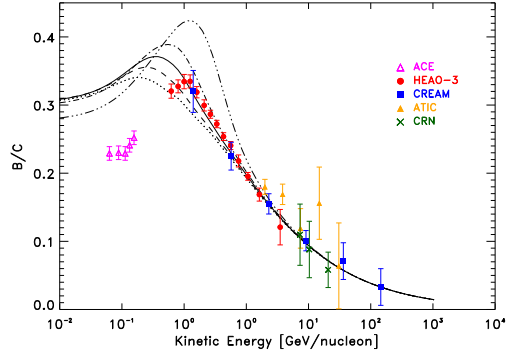}
   \label{fig:BC_variVA}
 }
  \subfigure[]
  {
    \includegraphics[scale=0.4]{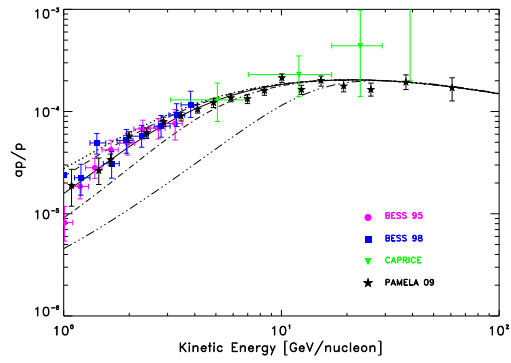}
    \label{fig:app_variVA}
 }
  \subfigure[]
  {
    \includegraphics[scale=0.4]{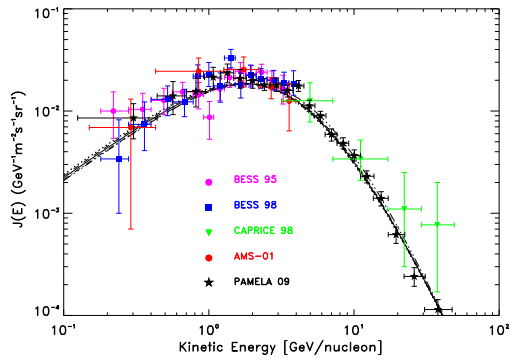}
    \label{fig:ap_variVA}
 }
  \subfigure[]
  {
    \includegraphics[scale=0.4]{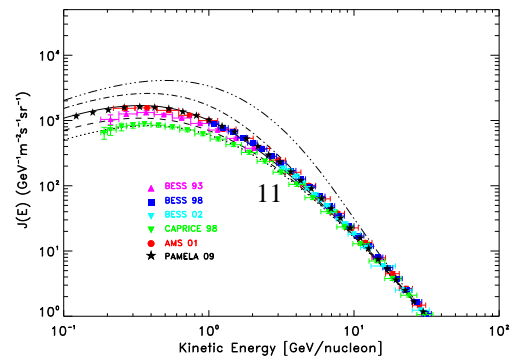}
    \label{fig:p_variVA}
 }
 \caption{The B/C (panel a), and the ${\bar p}/p$ (panel b) ratios, as well the antiproton (panel c) and proton (panel d) absolute spectra computed with DRAGON for $\delta = 0.45$ and $D_0/z_t = 0.8$ are plotted for several values of $v_A$ and compared with the respective experimental data. Dotted, short-dashed, solid, dot-dashed, long-dashed correspond to $v_A = 0,10,15,20,30~\km/\s$ respectively. %In panel (d) the oxygen spectrum, computed for $\delta = 0.45$,  $D_0/z_t = 0.8$ and $v_A = 15~\km/\s$, is compared with experimental data. 
 Here $\eta = 1$ which clearly does allow to match nuclear data below $1~\GeV/n$. For this reason the
modulation potentials $\Phi = 500~{\rm MV}$ adopted here for the B/C plot (as required to reproduce low energy Oxygen data) and $\Phi = 700~{\rm MV}$ for the ${\bar p}/p$  (to fit proton data) are not representative.  }
\end{figure}

Among those considered $v_A = 15~\km~ \s^{-1}$ is the  Alfv\`en velocity value which minimizes the $\chi^2$ of the combined analysis, hence it gives rise to the best overlap between the light nuclei and the ${\bar p}/{p}$ confidence regions.
This is also visible from Fig.s~\ref{fig:BC_variVA} and \ref{fig:app_variVA} where the B/C and the ${\bar p}/{p}$ ratios computed with $(\delta, D_0/z_t) \simeq (0.45, 0.7)$ are plotted for several $v_A$'s. 
It is also interesting to notice that the dependence of the ${\bar p}/p$ ratio on $v_A$ is driven by that of the proton spectrum since the absolute ${\bar p}$ spectrum is practically unaffected by re-acceleration (see Fig.\ref{fig:ap_variVA}).
We stress that Fig.s \ref{fig:BC_variVA},\ref{fig:p_variVA} are given here mainly for illustrative reasons since, below few GeV/n some other physics needs clearly to be introduced to reproduce the B/C data (see Sec. \ref{sec:le_model}).
  
Since the $v_A = 15~\km \s^{-1}$ combined analysis CL region is the largest, it also provides the most conservative constraints on $\delta$ and $D_0/z_{t}$. They are 
 $0. 3 \simleq \delta \simleq  0.6$ and $0.6 \simleq  D_0/z_{t} \simleq  1$ at  95\% CL. 
%Indeed, among the values considered in our analysis only $v_A = 15 ~\km/\s$ is allowed at $2 \sigma$. For this value of $v_A$ the 95\% CL allowed
%ranges of the other propagation parameters are $0.3 \simleq \delta \simleq  0.6$ and $0.63 < D_0/z_{t} < 0.73$ with best-fit at $(\delta, D_0/z_{t}) = (0.47, 0.76)$, practically coinciding with the result of the light nuclei analysis alone.  
%The overlap is even better if only PAMELA antiproton/proton data are considered (see below). 

 It should be kept in mind that our analysis accounts only for statistical experimental errors. Several systematic uncertainties, however, may affect our constraints too. Among them, systematic errors in the experimental data, uncertainties in the Galactic gas density and hydrogen fraction distributions and nuclear fragmentation cross sections play a major role.  A detailed discussion of the possible impact of these uncertainties on the determination of the CR propagation parameters is beyond the aims of this work. A thorough analysis was recently performed in  \cite{Maurin:09} showing that, if low energy data are accounted for (which requires to introduce several unknown parameters with respect to those considered in this work) the systematic uncertainties on the $D_0,~\delta$ and $v_A$ can be comparable, or even larger, than the statistical ones. 
However, the former uncertainties are significantly smaller if one considers only a subclass of models without convection and keeping fixed other parameters which only matters a low energies, as we do in this work.  For example, it was shown in  \cite{Maurin:09}  that for models with $v_c = 0$  the effect of considering different cross-section sets amounts to a $\sim 40\,\%$ uncertainty variation of $\delta$ which reduces to  $\sim 10\,\%$  if one considers only the most updated cross section sets. 

We verified with DRAGON that changing the GALPROP nuclear fragmentation cross sections with those given in \cite{Webber:03} produces only a marginal effect on the B/C ratio.  The relative effect of cross section uncertainties on the antiproton/proton ratio is negligible here due to the high statistical errors on those data.

%The fact that our data set includes measurements taken by different experiments in different conditions of solar activity might be relevant to the determination of our confidence regions. Moreover, since we do not account for charge dependent drift effects in the solar modulation, CR fluxes measured during phases having different polarities of solar magnetic field may be affected differently even if taken in low solar activity periods (see e.g. \cite{Bieber:1999dn} for a discussion about this issue).   
%While those kind of effects are expected to be small above 1 GeV/n, and to become irrelevant when rising $E_{\rm min}$ to higher value as we did in this work,  we decided to repeat our analysis by accounting for ${\bar p}{p}$ data taken from just one experiment, namely PAMELA \cite{Adriani:2008zq}. While the effect on the best fit parameters is tiny, as expected, 
%interestingly, the result of this further analysis is a strengthening of the constraints on the propagation parameters as the overlap between the light nuclei and  ${\bar p}/p$ confidence regions is even better in this case. Indeed, the combined analysis best fit values become $\left( \delta, D_0/z_t, v_A \right) = \left( 0.46, 0.76, 15 \right)$ 
%and the uncertainties on $(D_{0}/z_{t},\delta)$ reduce to $0.40 \simleq \delta \simleq 0.52$ and  $0.72 \simleq  D_0/z_t \simleq 0.82$.

\subsection{Maximal and minimal antiproton spectra}
\label{sec:max_min_models}

The previous results clearly favor a standard interpretation of the measured antiproton spectrum in terms of purely secondary production from CR nuclei. It is still possible, however, that a subdominant antiproton component arises from unconventional processes. 
In order to constrain such ``exotic" component(s) with experimental data, one has to compare antiproton data with the predictions of the theoretical models validated against CR nuclei data alone.  

For this purpose we define, for each value of $v_A$ considered in the above, a pair of MAX and MIN models which maximize and minimize respectively the antiproton absolute flux integrated in the range $1 - 100~\GeV$ under the condition to be compatible with secondary/primary light nuclei data down to $1~\GeV/{\rm n}$ within 95\% CL.  

\begin{figure}[tbp]
\centering 
\includegraphics[scale=0.4]{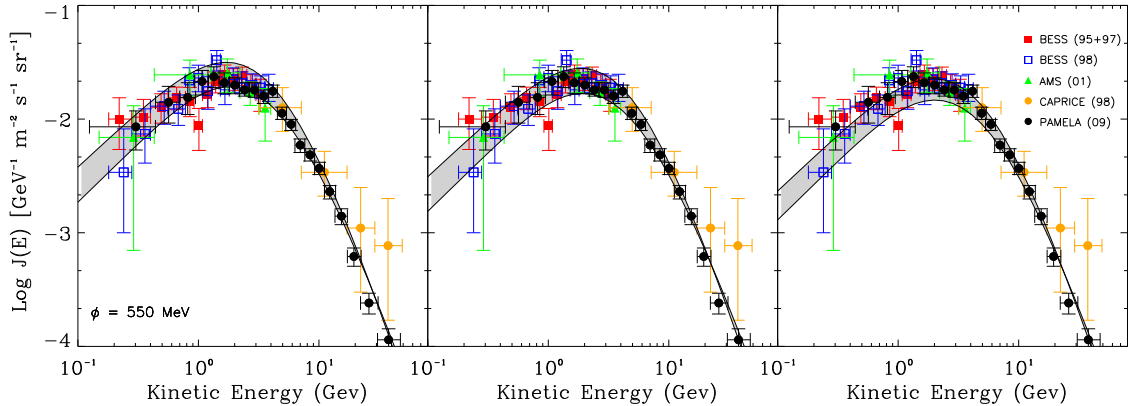}
\caption{The ${\bar p}$ absolute spectrum is shown for  $v_A = 10, 20, 30$ (from the left to the right panels respectively). The upper and lower curves correspond to the MAX and MIN models defined as in Sec.~\ref{sec:max_min_models} respectively.}
\label{fig:min_max}
\end{figure}

In Fig.~\ref{fig:min_max} we show the allowed ranges of the antiproton absolute spectrum for several values of $v_A$. Among the models considered here the absolute MAX and MIN models are those defined by the parameters $(\delta, D_0/z_t, v_A) =  (0.68, 0.46, 0)$ and $(0.30, 1.2, 30)$ respectively. 
Therefore, we conclude that, under the hypotheses adopted in this work, $\bar p$ constraints on an exotic component should not use, as propagation models, any model whose $\bar p$ background prediction is lower than our MIN (or larger than our MAX) model, as it would be in contrast with B/C data at 95\% CL. Hence, the most conservative constraint, under our hypotheses, arises from the request that the sum of the background $\bar p$ predicted by the MIN model plus the exotic $\bar p$ component do not exceed the experimental data, within some CL. %background levels for the CR $\bar p$ fluxes lower than our MIN model or larger than our MAX model should not be 
%Therefore, we conclude that, under the hypothesis adopted in this work, a possible unconventional component of antiproton cannot exceed the prediction of the absolute 
%MAX model without being incompatible with CR light nuclei measurements. {\bf Credo che questa conclusione sia falsa. La conclusione vera dovrebbe essere tipo: per fare constraints non e' possibile adottare un modello di BG superiore a MAX senza essere in conflitto con il B/C a 2 sigma}

 \section{A comprehensive model describing all data sets down to $0.1~\GeV/{\rm n}$}\label{sec:le_model}

The aim of this section is to test the consistency of our previous results with CR data below few GeV/n and to identify an effective model allowing to fit all available data. 
 
It is evident from Fig.~\ref{fig:BC_variVA} that while the best fit model obtained for $\eta = 1$ provides an excellent fit of experimental data above few $~\GeV/{\rm n}$,  below that energy it overshoots the B/C observations. 
As we discussed, such a discrepancy may be attributable to a number of effects which, at low energies, introduce degeneracies among the relevant parameters.
For this reason, a statistical analysis aimed to fit those low energy parameters against presently available data would be hardly interpretable (see e.g. \cite{Maurin:09}) and it is beyond the aims of this work. 

Here we follow a more phenomenological approach tuning only the parameter $\eta$ (see  Eq. \ref{eq:diff_coeff}) which sets the dependence of the diffusion coefficient on the particle velocity (a similar approach was followed in \cite{Maurin:09}).
Interestingly we find that the choice $\eta \simeq - 0.4$ allows to match light nuclei as well as antiproton data well below $1~\GeV/{\rm n}$ 
for almost the same range of  $\delta$ and $D_0/z_t$  values found for $\eta =1$.
Indeed, we checked that the $\eta = -0.4 $ and $\eta = 1$ CL regions computed for $E_{\rm min} =  5~\GeV/{\rm n}$  almost coincide (which is not the case for   $E_{\rm min} =  1~\GeV/{\rm n}$.
In Fig.s \ref{fig:BC_comp} - \ref{fig:protons}  we show as our best fit model obtained for  $\eta = - 0.4$,  $\delta = 0.5$, $D_0/z_t =  0.7$, and $v_A = 15~\km/\s^{-1}$ nicely reproduces all relevant data sets. They include also the N/O and C/O ratios (with $\sim 6~\%$ and $\sim 75~\%$ injection ratios respectively) as well as the absolute oxigen spectrum. 

We notice that the modified dependence of the diffusion coefficient upon rigidity, which is the consequence of adopting a value of $\eta$ different from $1$,  can be considered as an effective modelization of 
physics taking place at low energy, including some non-linear phenomena such as the dissipation of magneto-hydrodynamics (MHD) waves by their resonant interaction with CRs \cite{Ptuskin:2005ax}. Since this is the same interaction responsible for CR diffusion in the ISM, such an effect is unavoidable at some level. Interestingly, the value of $\delta$ used in \cite{Ptuskin:2005ax} to fit the B/C in the presence of MHD wave dissipation is  0.5, which is consistent with what we found here (differently from what we do here, however, a break in the injection index was invoked in that work). 

\section{Discussion and comparison with previous results}
\label{sec:discussion}

As we mentioned above, our numerical diffusion code DRAGON reproduces the same results of GALPROP \cite{GALPROPweb} under the same physical conditions. 
Our analysis and main conclusions, however, differ significantly from those reported in several papers based on that code.
  
\begin{figure}[tp]
  \centering
 \subfigure[]
 {
   \includegraphics[scale=0.35]{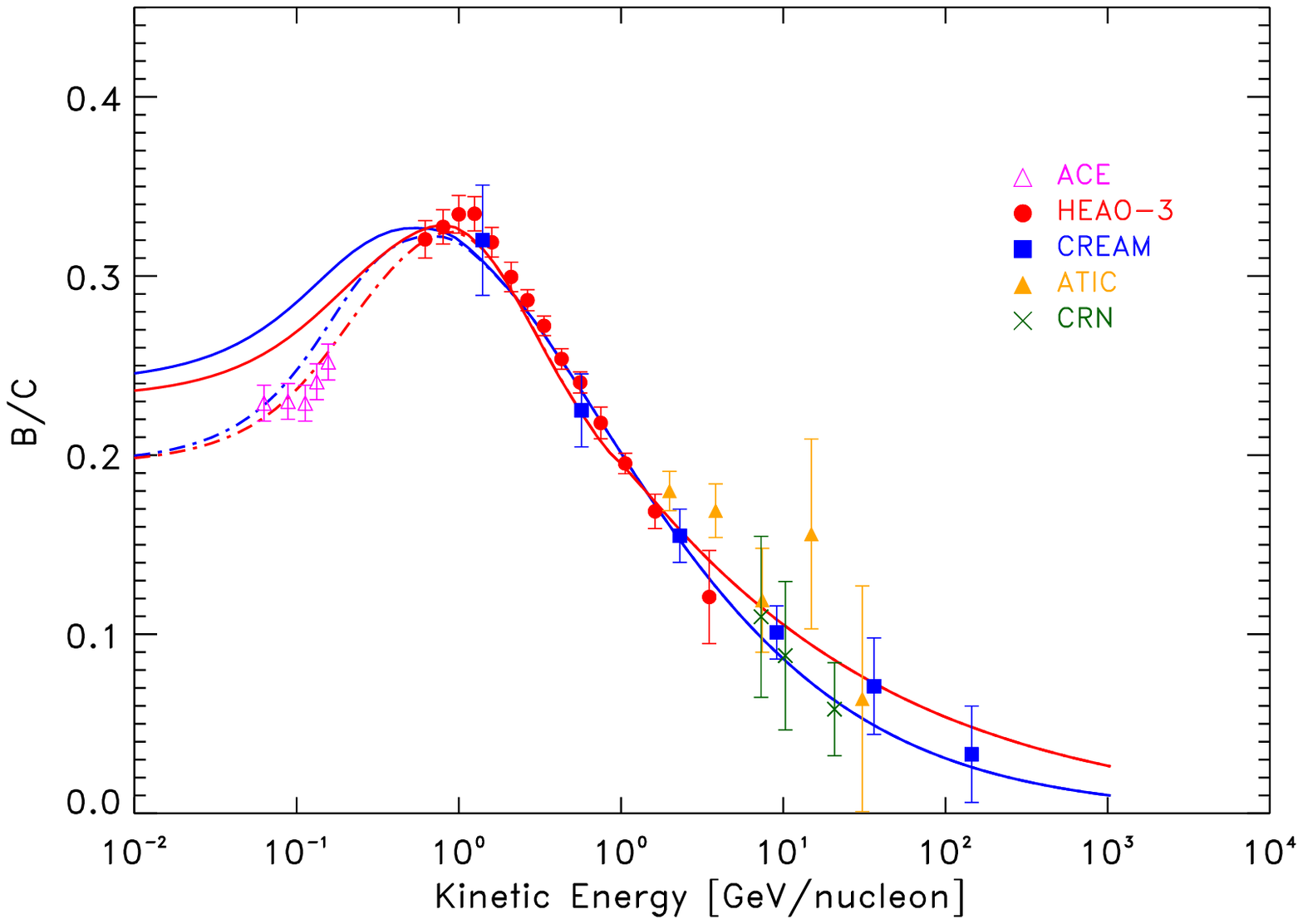}
   \label{fig:BC_comp}
 }
  \subfigure[]
 {
   \includegraphics[scale=0.35]{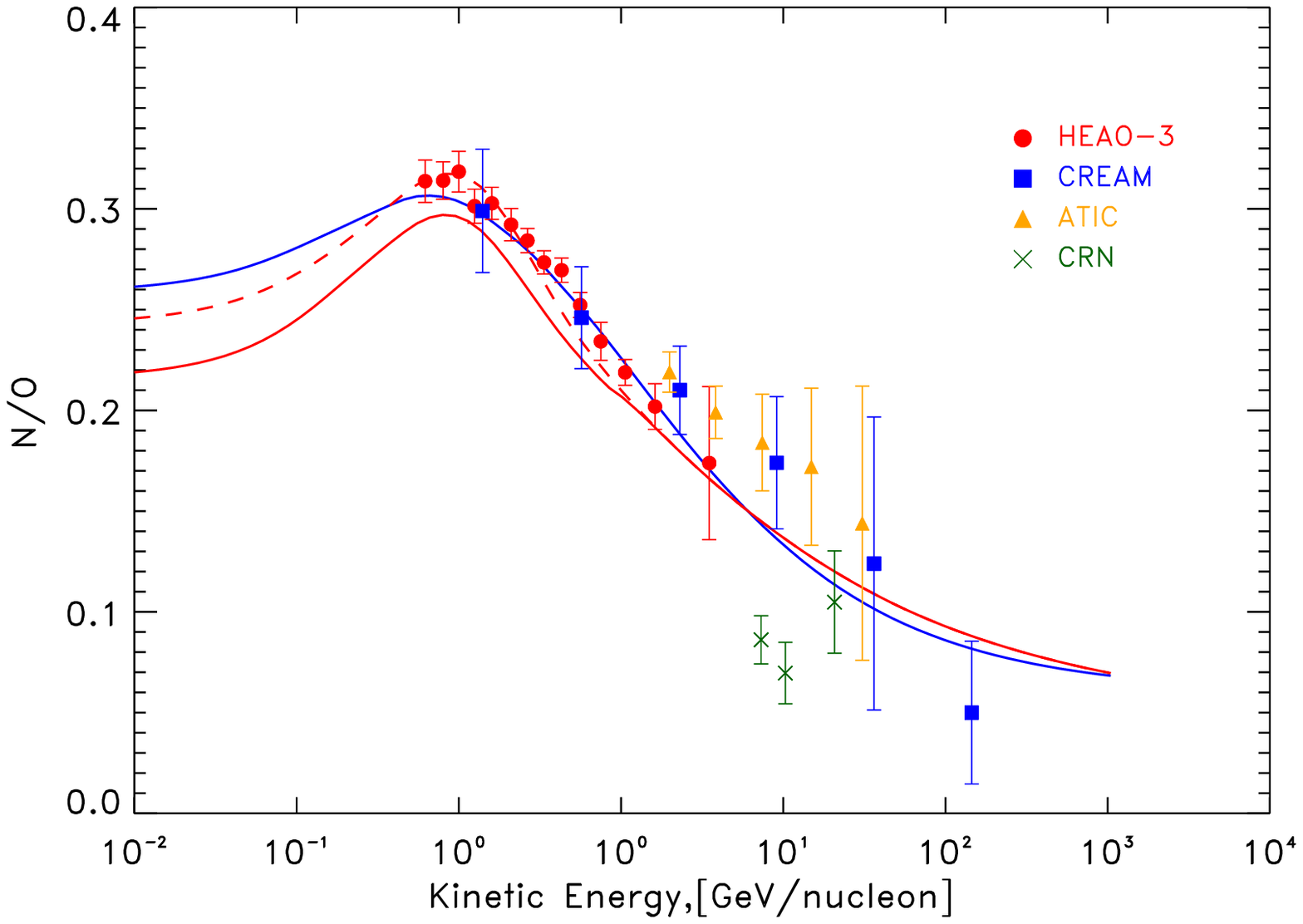}
   \label{fig:NO_comp}
 }
 \subfigure
 {
   \includegraphics[scale=0.35]{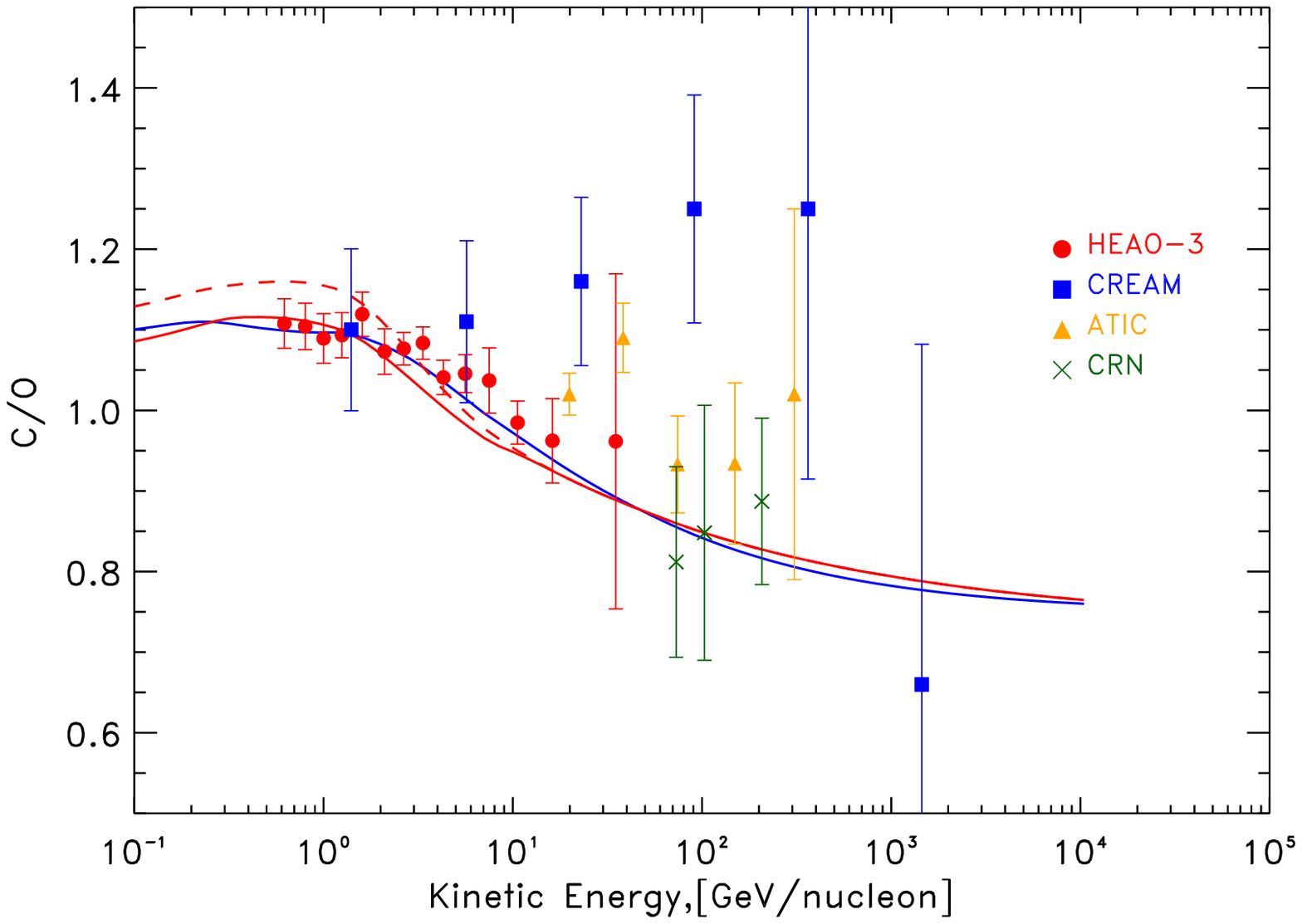}
   \label{fig:CO_comp}
  } 
 \subfigure
 {
   \includegraphics[scale=0.35]{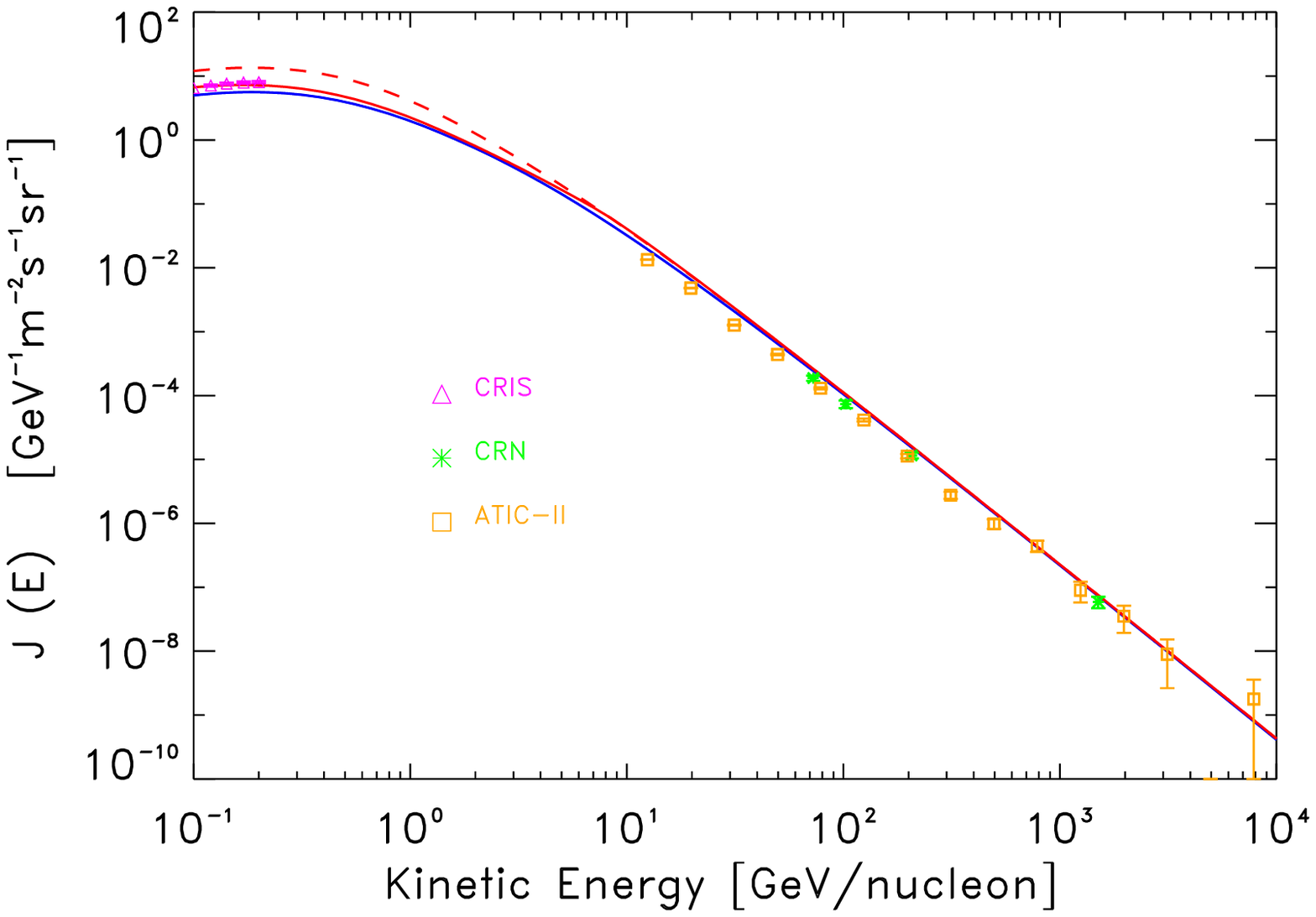}
   \label{fig:oxigen}
  }
\caption{The B/C (panel a), N/O  (panel b), C/O (panel c) and the oxigen absolute (panel d) spectra computed with our preferred {\it Kraichnan}
model (blue solid line), the  {\it Kolmogorov} reference model (red solid line) and the same model with no break in the CR source spectrum (red dashed line), are compared with available experimental data. In both cases we use DRAGON to model CR propagation and interactions (though almost identical results can be found with GALPROP). Here we use $\Phi = 450~{\rm MV}$ to modulate both the  {\it Kolmogorov model} and our {\it Kraichnan} reference models.  
$\Phi = 300~{\rm MV}$ was used only to match B/C ACE data which were taken in a very low activity solar phase.}
\end{figure}

\begin{figure}[tp]
  \centering
  \subfigure[]
  {
    \includegraphics[scale=0.35]{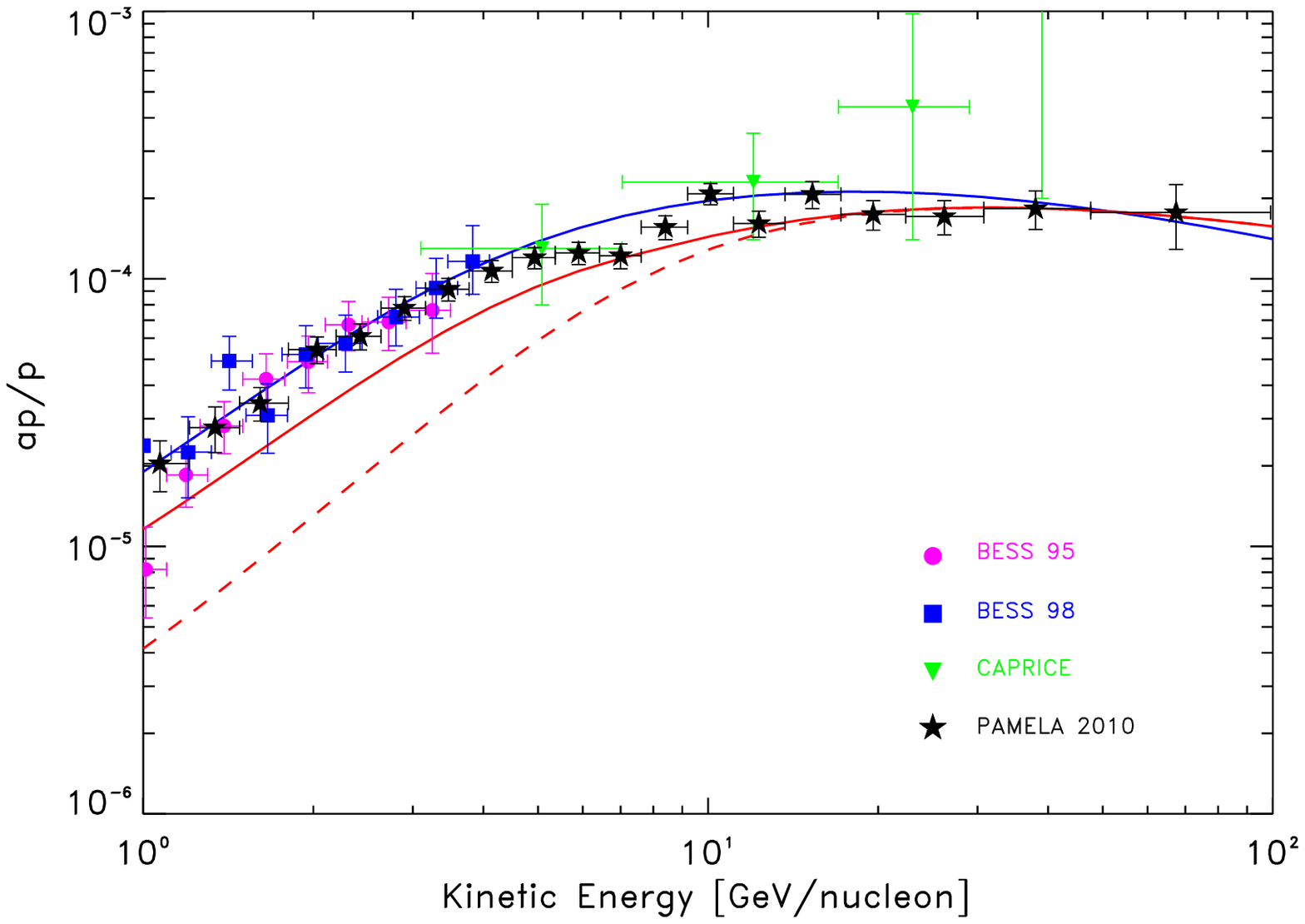}
    \label{fig:apratio_comp}
 }
   \subfigure[]
  {
    \includegraphics[scale=0.35]{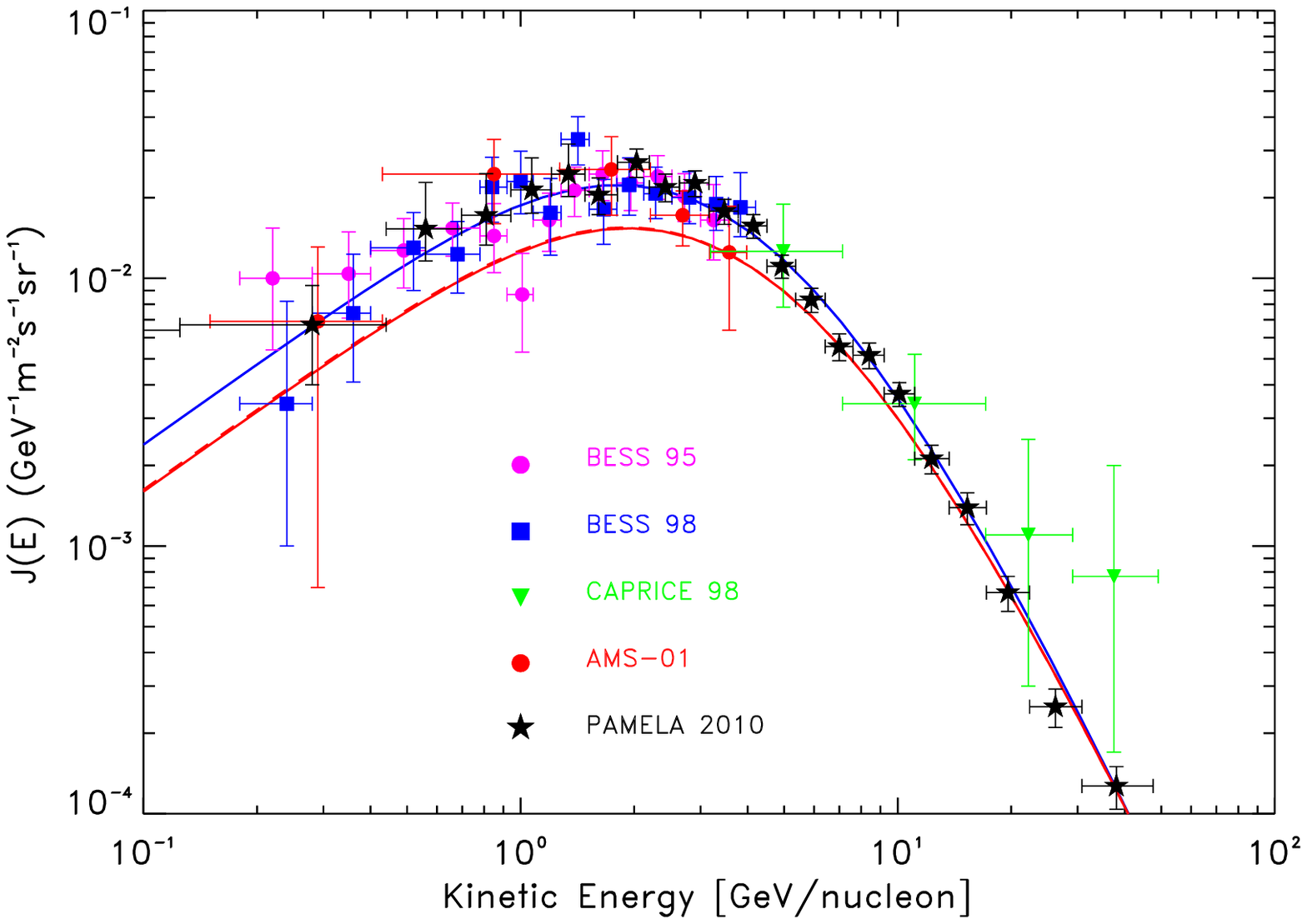}
    \label{fig:apflux_comp}
  }
   \subfigure[]
  {  
     \includegraphics[scale=0.35]{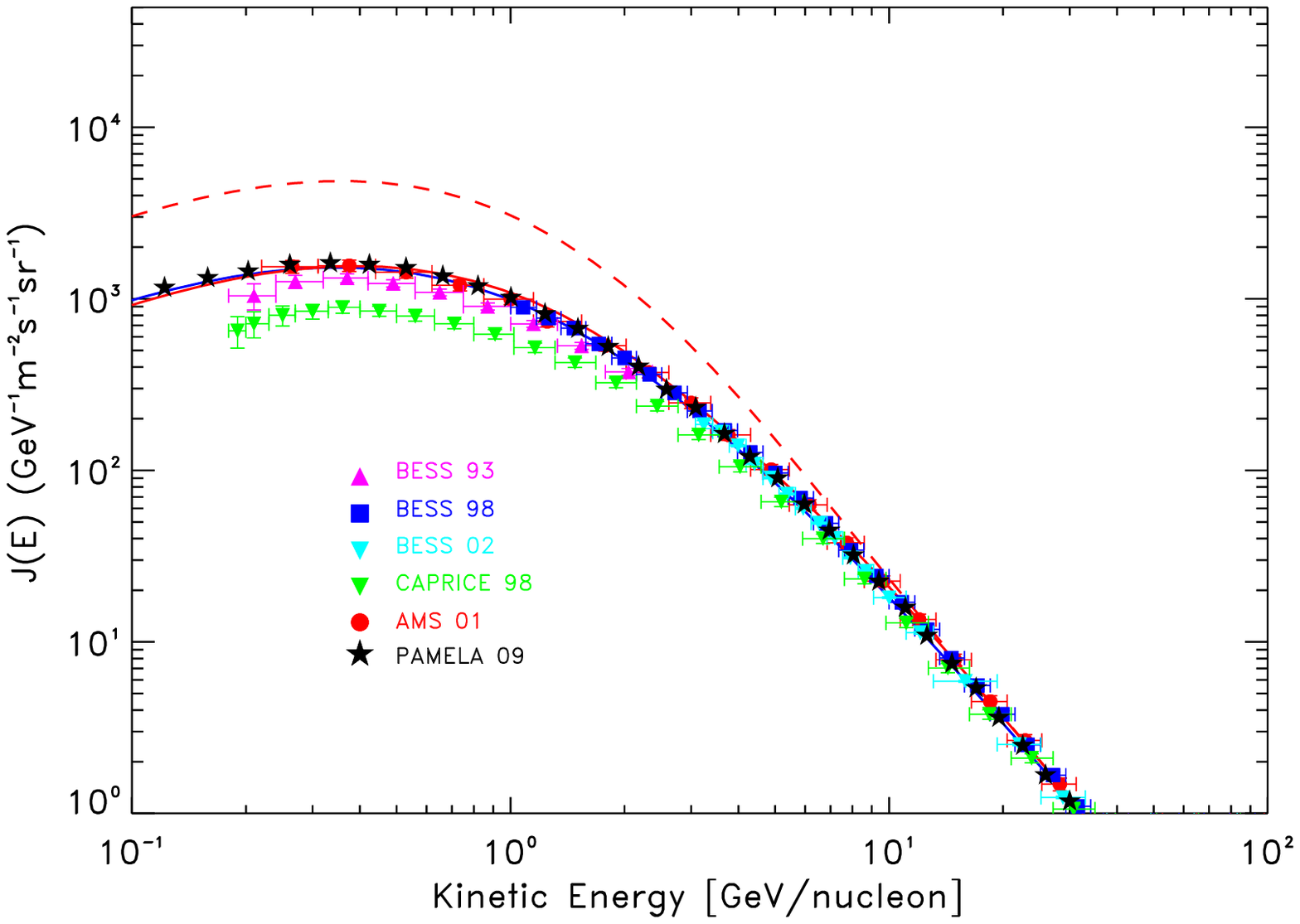}
      \label{fig:protons}
  }
\caption{The $\pbar/p$ (panel a), $\pbar$ (panel b) and proton (panel c) computed with DRAGON are reported here use the same models and line notation used in Fig.s 4. The solar modulation potential used here is $\Phi = 550~{\rm MV}$. In the (a) and (b) figures PAMELA 2010 data are taken from the recent paper \cite{pamela:2010rc} (see the note at the end of this paper).}
\end{figure}

In order to clarify the reasons of such a discrepancy, in Fig.~\ref{fig:BC_comp} and \ref{fig:apratio_comp} we compare the predictions of our reference diffusion-reacceleration model $(\delta, D_0/z_{t}, v_A, \eta) = (0.5, 0.7, 15, - 0.4)$, which for brevity we call {\it Kraichnan model}, 
with those obtained using the propagation parameters (and source distribution) of the {\it Kolmogorov model} discussed in \cite{Strong:04}, namely $(\delta, D_0(4~{\rm GV})/z_{t}, v_A, \eta) = (0.33,1.45, 30, 1)$ \footnote{In \cite{Strong:04} a spatially uniform diffusion coefficient ($z_{t}=z_{\rm max} = 4~\kpc$) was assumed. As we already noticed, for the purposes of the present analysis adopting a vertically uniform rather than varying diffusion coefficient only amounts to a rescaling of $D_{0}/z_{t}$. We verified that this does not affect any other result of our analysis.}.
%With respect to \cite{Strong:04} the $D_0/z_{t}$ have been rescaled here in order to account for the vertically uniform diffusion coefficient adopted for those models. 
 For the latter combination of parameters we consider two variants, represented by the solid/dashed red lines, which differ for the presence/absence of a break at $\rho_{\rm break} = 9~{\rm GV}$ in the CR nuclei source spectra. The {\it Kolmogorov model} considered in \cite{Strong:04} adopts such a break. It is evident from Fig. \ref{fig:protons} as this is needed in order to reproduce the low energy tail of the observed proton spectrum which otherwise could not be fit for any choice of the modulation potential.  It is important to notice, that this problem arises in all models with strong reacceleration $v_A > 20~\km~s^{-1}$. 
  On the other hand our Kraichnan reference model requires a ``modified"  behavior of the diffusion coefficient at low energy ($\eta = -0.4$ rather than $\eta = 1$) which, however, may be motivated by independent physical arguments as discussed in Sec.\ref{sec:le_model}.

From Fig.s \ref{fig:BC_comp} the reader can see that while both the Kraichnan and Kolmogorov models reproduce the B/C equally well, the former model provides a significantly better description of the N/O ratio measured by HEAO-3 \cite{HEAO-3} and CREAM \cite{CREAM} (see Fig.~\ref{fig:NO_comp}). Furthermore, what mostly favors our Kraichnan reference model are BESS \cite{Sanuki:2000wh}, CAPRICE \cite{Boezio:2001ac} and especially the PAMELA measurements of the ${\bar p}/p$ \cite{Adriani:2008zq} and antiproton absolute spectrum \cite{PAMELA:proton}. Indeed, the discrepancy between low energy antiproton data and the prediction of the ``conventional GALPROP model", which was already noted in \cite{Strong:04}, becomes more compelling due to the new PAMELA data, as shown in Fig.~\ref{fig:apflux_comp}.   

The comparison of our results with those of semi-analytical models is more difficult for obvious reasons. 
One of the difficulties lies in the simplified gas and source distribution adopted in those models (see Sec.~\ref{sec:code}). We verified, however, that such differences
only affect the constraints to $D_0/z_t$ with almost no effect on the determination of $\delta$. 
We also need to take into account that semi-analytical models (see e.g.~\cite{Maurin:01,Maurin:02}) assume diffusive reacceleration to take place only in the thin 
Galactic disk (whose height is $z_{d}$), while in the numerical models, as the one presented here, it takes place in the entire diffusion halo. Therefore, in order to compare the values of the Alfv\`en velocity in those papers with those reported in the above it is necessary to perform a proper rescaling. This is approximatively given by (see e.g.~Eq.~(18) in \cite{Maurin:2002ua}) $v_A = v_A^{\rm SA}\ \sqrt{z_d/z_t}$, with $v_A^{\rm SA}$ being the Alfv\`en velocity in the semi-analytical models and $z_t$
the half scale height of the Galactic disk. %It should also be noted that while in \cite{Maurin:01,Maurin:02} a statistical comparison between a large number of models and experimental B/C data was performed, in those works the C/O and N/O source ratios were not varied to be fitted against data. This is a significant simplification with respect to our approach for the reasons discussed at the beginning of Sec.~\ref{sec:nuclei_analysis}. Also the comparison of the theoretical antiproton spectrum with data was not treated statistically in \cite{Donato:01}.

In spite of these differences, and that CREAM and PAMELA data were not included in those analyses for chronological reasons, it is comforting that for low values of the convective velocity $v_c \simeq 0$ the preferred value of $\delta$ estimated in \cite{Maurin:01,Maurin:02} is in remarkably good agreement with that found in this work: $\delta \simeq 0.45$. Interestingly, the rescaled value of $v_A$ determined in \cite{Maurin:01} is $v_A \simeq 10$, for $v_c \simeq 0$, which is also in good agreement with our results.  It is important to notice that, similarly to what we did in our analysis, no break in the source spectral index was assumed in \cite{Maurin:01,Maurin:02}.

We remind the reader that in the above we always assumed $v_c = 0$ as higher values of that parameter are not required to interpret CR nuclei and antiproton data. 
Models with a finite $v_c$, which may also allow to fit low energy data though with a different combination of some parameters ($\eta$, the modulation potential $\Phi$, or any other), will be considered elsewhere. We already tested, however,  that taking $v_c$ in a reasonable range of values do not affect significantly our constraints on most relevant diffusion coefficient parameters, namely $\delta$ and $v_A$.      
Indeed, we verified that for various choices of the convective velocity, and of its vertical gradient, it is always possible to rescale the
diffusion coefficient normalization $D_0/z_t$ so that both the B/C and the antiproton-to-proton ratio remain almost unaffected above $5~\GeV/{\rm n}$, i.e. the energy range we considered in our analysis. Our tests confirm what is claimed in \cite{Strong:98}: namely, that the contribution of convection to the B/C energy slope is negligible, especially in the intermediate and high energy regions.

\section{Conclusions}
\label{sec:conclusions}

We used recent data on CR light nuclei and antiprotons to determine the conditions of propagation of high energy CRs in the Galaxy, exploiting our numerical code, DRAGON. In the framework of a diffusion-reacceleration model, we performed a thorough analysis of the agreement of our predictions with experimental information, aimed at constraining, in a statistical sense, the most important model parameters: $D_{0}/z_{t}$, $\delta$ and $v_{A}$. The amount and quality of data is enough to allow us to perform our analysis in a wide energy range, from 1 to $10^3~\GeV/{\rm n}$, and also to check the evolution of our results varying the minimal energy at which data are considered. This is essential to reduce the uncertainties related to possibly unknown low energy physics, including solar modulation, and to disentangle the effects of reacceleration from those of diffusion.

One of the most important results of this analysis is that light nuclei (especially B/C) data and antiproton data can fit into a unique, coherent diffusion-reacceleration model of propagation, as it can be read off Fig.~\ref{fig:CL}. Indeed, the confidence regions obtained for $E > 5~\GeV/{\rm n}$ (where only the effects of diffusion and re-acceleration matters),  light nuclei and antiproton CL regions nicely overlap to produce combined constraints on $D_{0}/z_{t}$, $\delta$.  
 %The agreement becomes even better if only PAMELA ${\bar p}{p}$ are considered.
While this was also shown in previous works (which however did not exploit the new CREAM data), a combined statistical analysis of nuclear and antiproton data has been performed here for the first time. We showed that such an analysis allows to narrow significantly the allowed values of $\delta$ and $D_{0}/z_{t}$: our constraints  $0.3 < \delta < 0.6$ and $0.6 < D_{0}/z_{t} < 1$, as obtained at 95\% C.L., are significantly more stringent that those previously determined in the related literature.  
Furthermore we found, for the first time,  that only a relatively narrow range ($10 - 20~\km/\s$) of the Alfv\`en velocity values are allowed.
Even well below $5~\GeV/{\rm n}$, we showed that it is possible to find effective models which, still fulfilling those constraints, allow to nicely reproduce all relevant data.
  
%In particular, only $v_{A}\simeq 15~\km/\s$ is allowed. For this value of the Alfv\`en velocity, the range of the other parameters, allowed at 95\% CL, are $0.38 < \delta < 0.57$ and $0.63 < D_{0}/z_{t} < 0.73$, with best-fit at $(\delta, D_0/z_{t}) = (0.47, 0.76)$ (we remind the reader that $D_0/z_{t}$ is expressed here in units of $10^{28}~\cm^2~\s^{-1}~\kpc^{-1}$ and $v_A$ in $\km~\s^{-1}$).   

We also found that the preferred values of the N/O and C/O ratios at injection are $\sim 6~\%$ and $\sim 75~\%$ respectively. These results, and in particular the analysis of data with $E_{\rm min} = 5~\GeV/{\rm n}$, clearly favor a Kraichnan like CR diffusion ($\delta = 0.5$) respect to Kolmogorov ($\delta = 0.33$).  It is worth noticing that a relatively large value of $\delta$, as that preferred by our analysis, would give rise to a too large CR anisotropy if our results are extrapolated to $E_k \gg 10^{5}~\GeV/{\rm n}$ (see e.g.~\cite{Blasi:2008ch} and ref.s therein). Our results, therefore, may call for some changes in the standard CR propagation scenario. 

While the effects of systematic uncertainties on fragmentation cross-sections
are not studied in details in this work,  recent results suggests that they should be smaller than the quoted $2 \sigma$ statistical uncertainties on the transport parameters. Indeed we performed DRAGON runs using  a different set of nuclear cross-sections as determined in \cite{Webber:03}, finding the same B/C and N/O ratio found in the above within few \%'s.   

Given that anyway nuclei data alone are able to provide constraints on $D_{0}/z_{t}$ and $\delta$, we use this information to establish a range for the maximal and minimal flux of antiprotons expected from CR interactions in the gas and still compatible with light nuclei observations within 95\% CL. This range information can be used as a CR background in analyses aimed at constraining or finding some exotic signal in antiproton data.

Forthcoming data from several running or scheduled experiments, as PAMELA (both for antiprotons and light nuclei), CREAM-II \cite{CREAM3}, TRACER \cite{Boyle:2008ut,Ave:2008uw}, and AMS-02 \cite{ams02} which will measure both CR nuclei and ${\bar p}$ fluxes from hundreds MeV/n up to TeV/n, will soon allow tighter constraints.  
Especially AMS-02 is expected to provide very accurate data and, what is most relevant here, it will allow simultaneous and consistently calibrated measurements of several nuclear species 
and antiprotons (as well as electrons and positrons which will also provide valuable complementary inputs).
The AMS-02 potential to pinpoint the CR propagation was also recently showed in \cite{Pato:2010ih} where, however, the power of a combined analysis of CR nuclei and antiproton data was not discussed.

\section*{Note added}

When this paper was in its final refereeing process, PAMELA collaboration published updated data of the ${\bar p}/{p}$ ratio and ${\bar p}$ absolute spectrum \cite{pamela:2010rc}. The  ${\bar p}/{p}$ data differ very little from those we used in our statistical analysis (discussed in Sec. 3) which were taken from \cite{Adriani:2008zq} (PAMELA 2009). Therefore, their update should not affect significantly our constraints on the propagation parameters. In Fig. \ref{fig:apratio_comp} and \ref{fig:apflux_comp} we display the new PAMELA antiproton data. It is evident as our best fit model, which was determined using ${\bar p}$ PAMELA 2009 data, nicely matches also PAMELA 2010. 

\section*{Acknowledgments}
We are indebted with P.~Ullio for invaluable comments and suggestions. We warmly thank P.~Picozza for allowing us to extract preliminary PAMELA proton and antiproton data from his talk at TeVPA 2009. We thank F.~Donato, P.~Maestro, G.~Sigl and A.~W.~Strong for reading the draft of this paper and providing useful comments. We also thank D.~Maurin for giving us the electronic form of the nuclear cross section tables reported in \cite{Webber:03} under W.R.~Webber kind permission.

D.~Grasso is supported by the Italian Space Agency under the contract AMS-02.ASI/AMS-02 n.~I/035/07/0.  
D.~Grasso and D.~Gaggero acknowledge partial financial support from UniverseNet EU Network under contract n.~MRTN-CT-2006-035863.  LM acknowledges support from the State of Hamburg, through the Collaborative Research program ``Connecting Particles with the Cosmos'' within the framework of the LandesExzellenzInitiative (LEXI).

%\section*{References}

\end{document}